\documentclass[iop,apj,tighten,twocolappendix,numberedappendix]{emulateapj}
\usepackage{apjfonts}

\usepackage{graphicx}
\usepackage{amsmath}
\usepackage{amssymb}
\usepackage{color}
\usepackage{mathtools}
\usepackage{url}

\usepackage[breaklinks,colorlinks,citecolor=blue,linkcolor=red]{hyperref} 
\usepackage[all]{hypcap}

\newcommand\msun{\, \rm M_\odot}

\newcommand\kms{\, \rm km\,s^{-1}}

\newcommand\gpcyr{\, \rm Gpc^{-3}\,yr^{-1}}

\newcommand\mcrit{{m_{\rm crit}}}

\begin{document}

\title{Demographics of neutron stars in young massive and open clusters}

\author{Giacomo Fragione\altaffilmark{1,2}, Sambaran Banerjee\altaffilmark{3,4}}
\affil{$^1$Department of Physics \& Astronomy, Northwestern University, Evanston, IL 60208, USA} 
\affil{$^2$Center for Interdisciplinary Exploration \& Research in Astrophysics (CIERA)}
\affil{$^3$Helmholtz-Instituts f\"{u}r Strahlen- und Kernphysik (HISKP), Nussallee 14-16, D-53115 Bonn, Germany}
\affil{$^4$Argelander-Institut f\"{u}r Astronomie (AIfA), Auf dem H\"{u}gel 71, D-53121 Bonn, Germany}

\begin{abstract}
Star clusters appear to be the ideal environment for the assembly of neutron star-neutron star (NS-NS) and black hole-neutron star (BH-NS) binaries. These binaries are among the most interesting astrophysical objects, being potential sources of gravitational waves (GWs) and gamma-ray bursts. We use for the first time high-precision N-body simulations of young massive and open clusters to study the origin and dynamical evolution of NSs, within clusters with different initial masses, metallicities, primordial binary fractions, and prescriptions for the compact object natal kicks at birth. We find that the radial profile of NSs is shaped by the BH content of the cluster, which partially quenches the NS segregation due to the BH-burning process. This leaves most of the NSs out of the densest cluster regions, where NS-NS and BH-NS binaries could potentially form. Due to a large velocity kick that they receive at birth, most of the NSs escape the host clusters, with the bulk of their retained population made up of NSs of $\sim 1.3\msun$ coming from the electron-capture supernova process. The details of the primordial binary fraction and pairing can smear out this trend. Finally, we find that a subset of our models produce NS-NS mergers, leading to a rate of $\sim 0.01$--$0.1\gpcyr$ in the local Universe, and compute an upper limit of $\sim 3\times 10^{-2}$--$3\times 10^{-3}\gpcyr$ for the BH-NS merger rate. Our estimates are several orders of magnitude smaller than the current empirical merger rate from LIGO/Virgo, in agreement with the recent rate estimates for old globular clusters.
\end{abstract}


\section{Introduction}
\label{sect:intro}
Neutron stars (NSs) are the end product of the evolution of stars of moderate mass ($\sim 7\msun$--$20\msun$, depending on the metallicity). About 20 NS-NS binaries have been observed in the radio band in our Milky Way Galaxy alone \citep[e.g.,][]{martinez2017pulsar,tauris2017formation,cameron2018high,lynch2018green,stovall2018palfa,Ridolfi2019upgraded}, some of which might have been formed in, and then ejected from, dense star clusters \citep{andrews2019double}. Gravitational wave (GW) detectors promise to observe hundreds of merging NS-NS and BH-NS (black hole-neutron star) binaries. Two NS-NS binaries have been observed so far by the LIGO/Virgo collaboration, including the most massive NS-NS binary ever observed \citep{abbott2017gw170817,abb2020}. The merger rate in the local Universe has been estimated $\sim 250$--$2810\gpcyr$. No BH-NS binaries have been confirmed either through radio observations or GW emission\footnote{Possible candidates can be found at \url{https://gracedb.ligo.org/}.}, with a LIGO/Virgo $90\%$ upper limit of $\sim 610\gpcyr$ on the merger rate \citep{abb2019cat}. The recent discovery of GW190814, a coalescence of a $\sim 23\msun$ BH with a $\sim 2.6\msun$ compact object, could be the either be a binary BH merger (with a BH in the mass gap) or the first detected BH-NS merger, with a rate of $\sim 1$--$23\gpcyr$ \citep{abbo2020}. In any case, NS-NS and BH-NS mergers result of a particular interest since they are followed by a short gamma-ray burst, which can reveal the details of the explosion mechanism and neutron star structure \citep[e.g.,][]{narayan1992grb,berger2014short,dora2016}. 

Several different astrophysical channels have been proposed to form merging compact objects. Possibilities include isolated binary evolution through a common envelope phase \citep{bel16b,kruc2018} or through chemically homogeneous evolution \citep{demink2016,march16}, mergers in star clusters \citep{askar17,baner18,frak18,rod18,kremer2020}, Kozai-Lidov (KL) mergers of binaries in galactic nuclei \citep{antoper12,petr17,fragrish2018,grish18}, in stellar triple \citep{ant17,sil17,frl2019a,frl2019b,fragetal2020} and quadruple systems \citep{fragk2019,liu2019}, GW capture events in galactic nuclei \citep{olea09,rass2019}, and mergers in active galactic nuclei accretion disks \citep{bart17,sto17,secunda2019}. Each model can potentially account for a significant fraction of the overall observed population, predicting a rate of $\sim\ 1$--$100\gpcyr$ in the local Universe. With hundreds of confirmed events from future detections, the statistical contribution of different channels could be disentangled by using the distributions of their masses, spins, eccentricities, and redshifts \citep{olea16,gondan2018}.

In the dense environment of star clusters, the frequent three- and four-body encounters can efficiently drive binary BHs to merger. A number of studies have shown that merging BH--BH binaries are formed at high enough rates in star clusters to potentially explain the LIGO/Virgo detection rate \citep[e.g.,][]{fragione2018black,rodriguez2018posta,rodriguez2018postb,samsing2018black,Choksi2019star,kremer2019post,samsing2019gravitational}.

Recently, the possible contribution of star clusters to NS--NS and BH--NS mergers has become under scrutiny. \citet{ye2020} estimated that the merger rate of NS--NS and BH--NS binaries in globular clusters is $\sim 0.02$--$0.04\gpcyr$, about $\sim 3$--$4$ orders of magnitude smaller than the inferred LIGO/Virgo rates. The incapacity of catalyzing these mergers has to be ascribed to the BH-burning process \citep{Kremer2019d}, where stellar BHs dominate the cluster cores and prevent the mass segregation of NSs \citep{frag2018,ye2019millisecond}. As a result, the rates of dynamical interactions involving NSs are reduced and merging binaries containing NSs cannot be efficiently formed. A similar result was found by \cite{belczynski2018origin}, who derived a NS--NS merger rate from globular clusters of $\sim 0.05\gpcyr$. Similar results were obtained by \citet{bae2014compact}, \citet{clausen2013black}, and \citet{asedda2020} for globular and nuclear star clusters.

In this paper, we use for the first time high-precision self-consistent N-body simulations of young massive and open clusters to study the origin and dynamical evolution of NSs. Our direct $N$-body simulations are performed with the state-of-the-art collisional evolution code \textsc{Nbody7} \citep{aseth2003,aseth2012}, with the most up-to-date prescriptions for single and binary stellar evolution \cite{baner2019bse}. To this end, we use a suite of  a set of $65$ $N$-body simulations, presented for the first time in \citet{banerjee2020}, encompassing different cluster masses, initial fractions of primordial binaries, and metallicities. For the first time, our findings show that also young massive and open clusters are not an efficient factory of BH-NS and NS-NS mergers, producing a merger rate about $\sim 3$--$4$ orders of magnitude smaller than the LIGO/Virgo rates.

Our paper is organized as follows. In Section \ref{sect:models}, we describe our numerical models of dense star clusters. In Section \ref{sect:masseg}, we discuss the role of the mass segregation and black hole heating. In Section \ref{sect:binret}, we discuss the role of the primordial binary fraction and pairing on the retention of neutron stars, while, in Section~\ref{sect:nsbinary}, we present the population of compact object binaries containing neutron stars, that we find in our simulations. In Section \ref{sect:conc}, we summarize our findings and draw our conclusions.

\section{N-body models}
\label{sect:models}

We utilize direct $N$-body evolutionary models of star clusters, computed using the most up-to-date version of \textsc{Nbody7} \citep{aseth2012}. The updates include the treatment of stellar winds, the prescriptions for the natal kicks imparted to remnants at formation and the natal BH spins, the inclusions of relativistic recoil kicks as a result of BH mergers, and the prescriptions for star-star and star-remnant mergers. For details see \citet{banerjee2020}.

The cluster models we consider represent young massive and open clusters that continue to form and dissolve throughout gas-rich galaxies, as in our Galaxy and the Local Group. The star clusters we model have initial sizes of the order of a parsec, consistent with gas-free young clusters in the Milky Way and neighbouring galaxies \citep{portgz2010,banerjkroupa2017} \footnote{These clusters are assumed to have survived their assembling and violent-relaxation phases, and have expanded to parsec scale sizes from sub-parsec sizes, as observed in newly-formed, gas-embedded, and partially-embedded clusters and associations \citep[e.g.,][]{bankroupa2018}.}.

The initial model clusters have \citet{plummer1911} profiles with masses $1.0\times 10^4\msun$--$1.0\times 10^5\msun$, half-mass radii $1.0$ pc--$3.0$ pc, and metallicities $0.0001$--$0.02$. All the computed models are taken to be initially in dynamical equilibrium and unsegregated, subjected to a solar-neighbourhood-like external galactic field \citep[see Table~C1 in][]{banerjee2020}.

Initial stellar masses are sampled from a \citet{kroupa2001} initial mass function, in the range $0.08\msun$--$150\msun$. The overall primordial binary fraction in our models is set to $0.0$, $0.05$, $0.10$. The initial binary fraction of the OB-type stars, of ZAMS mass $m_{\rm ZAMS}\ge \mcrit=16\msun$, which is taken separately, is $\sim 100\,\%$ \citep{baner18}, to be consistent with the observed high binary fraction among the OB-type stars in young clusters and associations \citep[see, e.g.,][]{sanaevans2011,moedist2017}. In two of the models, we vary the critical mass and fix it to $\mcrit=5\msun$ \citep[see][]{banerjee2020}. For $m_{\rm ZAMS}\ge\mcrit$, we pair an OB-star only with another OB-star, as it is typically observed, and the pairing among the lower mass stars is obtained separately. The $m_{\rm ZAMS}\ge\mcrit$ binaries are taken to initially follow the orbital-period distribution of \citet{sanaevans2011} and a uniform mass-ratio distribution. The orbital periods of the lower mass primordial binaries follow the \citet{duq1991} distribution and their mass-ratio distribution is also taken to be uniform. The initial binary eccentricities are drawn from a thermal distribution \citep{spitzer1987} for the binaries with components $m_{\rm ZAMS}<\mcrit$ and from the \citet{sanaevans2011} distribution for the $m_{\rm ZAMS}\ge\mcrit$ binaries.

For what concerns stellar evolution and the formation of compact objects, we consider both pair instability and pulsation pair instability supernovae for BH formation in our simulations \citep{bel2016b}. Moreover, we include models where BHs and NSs are born as a result of rapid and delayed supernova \citep[SN;][]{fryer2012}. In our simulations, the maximum possible NS mass is about $2.5\msun$. The amount and fraction of the supernova material fallback are provided by the chosen remnant-mass scheme. The remnant natal kick is taken to be Maxwellian with mean velocity $\sigma \sim 265\kms$, based on observed kick distribution of Galactic NSs \citep{hobbs2005}. However, NSs that are born in the electron-capture supernova (ECS) are assumed to have a natal kick of the order of $\sim 5\kms$ \citep{pod2004}. BH natal kicks are assigned either assuming momentum conservation \citep{fryerkalo2001} or collapse asymmetry \citep{burrows1996,fryer2004}. For full details see \citet{baner2019bse} and \citet{banerjee2020}.

All models are evolved until $11$ Gyr, unless the cluster is dissolved earlier \footnote{The present model clusters initiate and remain over most of their evolution well tidally under-filled, so that they have long dissolution times, $>11$ Gyr for most of them  \citep[see]{Banerjee2017stellara}. Therefore, they continue to exhibit significant dynamical activities for long evolutionary times which is why most models are run up to $11$ Gyr \citep{banerjee2020}.}.

\begin{figure*} 
\centering
\includegraphics[scale=0.85]{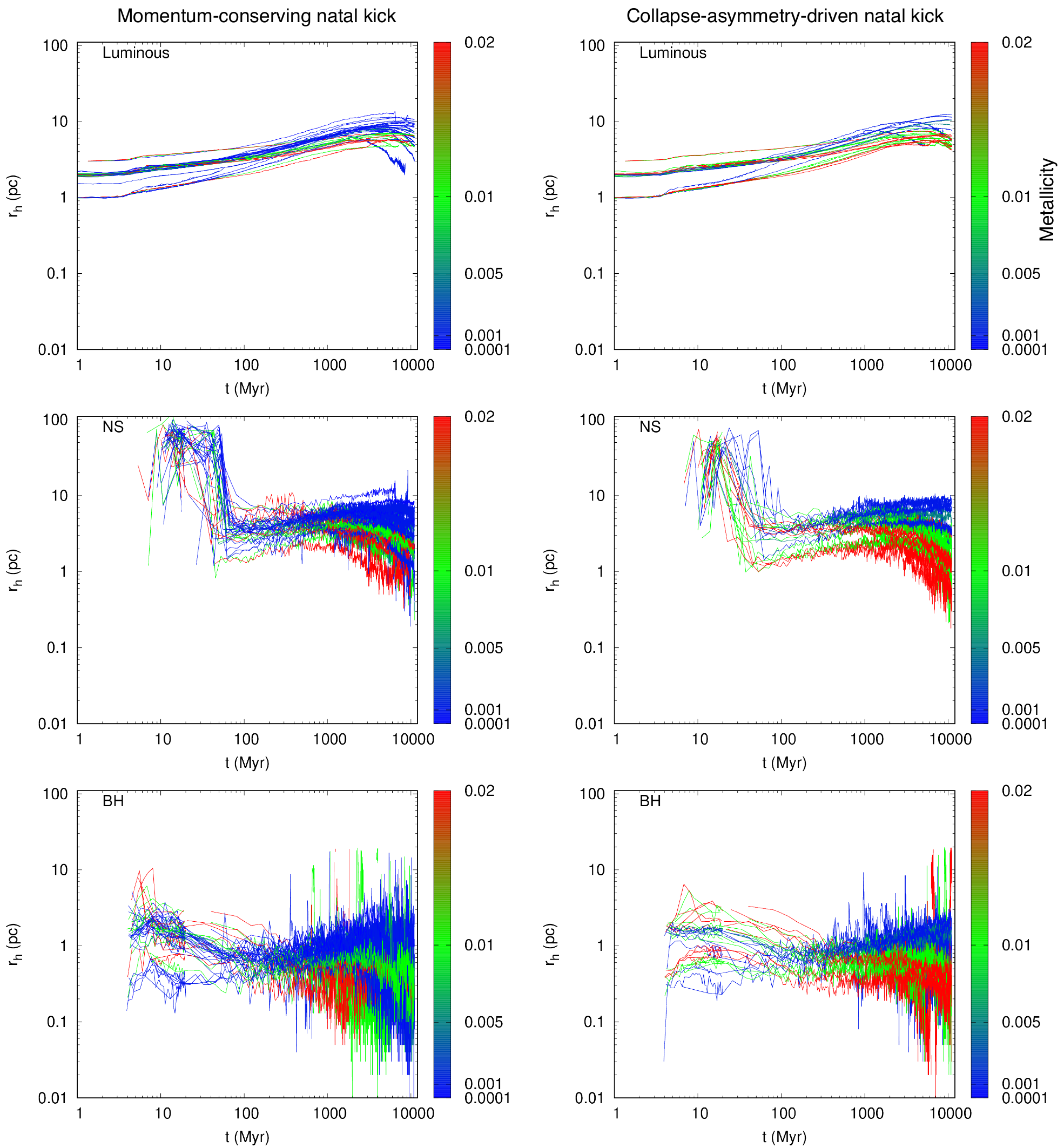}
\caption{Half-mass radius distance from the center of the host cluster as a function of time of luminous stars (top panel), neutron stars (central panel), and black holes (bottom panel), for the cluster models presented in \citet{banerjee2020}. Left panel: cluster models assuming momentum-conserving natal kicks; right panel: cluster models assuming collapse-asymmetry-driven natal kicks. Color code: cluster metallicity.}
\label{fig:rhevol}
\end{figure*}

\section{Neutron star segregation and black hole heating}
\label{sect:masseg}

NSs form from the collapse of massive stars within the first $\sim 100$\,Myr of the cluster lifetime. In low-metallicity environments, both the lower and the upper ZAMS (zero-age main sequence) limits of NS formation are smaller than those at high metallicities, thus leading to NS formation over longer evolutionary times. The NSs that form as a result of a core-collapse SN leave the host cluster shortly after their birth, since the cluster escape speed is typically much lower than the natal kick velocity they receive. Only in nuclear star clusters, where the escape velocity is of the same order as $\sigma$, a non-negligible fraction of core-collapse NSs could remain bound \citep[see e.g.,][]{Banerjee2017stellara,baner18,banerj2018s}. The bulk of the NS population that remains bound to the host star cluster is mainly produced via ECS, since they receive a small kick at birth.

Although the retained NSs are, on average, heavier and remain more concentrated than regular stars, their mass segregation is also in part quenched due to the BH-burning process \citep{frag2018,Kremer2019d}. Initially, when the BH-engine is active, strong dynamical encounters among the BHs act as a energy source for the rest of the cluster. Moreover, when a large BH population remains in a cluster, the cluster exhibits a large observed core radius \citep{Kremer_2018}. For lower metallicities, the BHs could become more massive and numerous, resulting in higher energy deposition in the rest of the cluster members. As the BH population erodes and the engine is weakened, the cluster core radius shrinks and the NSs start efficiently segregate towards the densest regions of the cluster. If the BHs are almost fully depleted, the host cluster attain a structure that would observationally be identified as core-collapsed, with the NSs occupying the central regions \citep{Kremer_2018,ye2020}.

Figure~\ref{fig:rhevol} illustrates the evolution of the half-mass radius distance from the center of the host cluster as a function of time for luminous stars (top panel), NSs (central panel), and BHs (bottom panel), for the cluster models presented in \citet{banerjee2020}. The spikes in the BH distributions come from slow escaping BHs due to evaporation or slow ejections as a consequence of dynamical interactions with other BHs. The spikes in the NS panels are from fluctuations due to the small number of remaining objects close to the cluster's dissolution in a few of the models, typically, among the lowest metallicity and least dense models. We show both cluster models where momentum-conserving natal kicks (left panel) and collapse-asymmetry-driven natal kicks (right panel) are assumed. We find that there is no a significant difference between these two cases, that is the adopted scheme for natal kicks do not affect the dynamical behavior of the NS and BH population.

Roughly independently from the cluster model, Figure~\ref{fig:rhevol} shows that the distribution of the NS population typically overlaps with the stellar population. However, NSs do not overlap significantly with the BH population in the innermost cluster regions, which are the most favorable place to form NS-NS and BH-NS binaries owing to their high densities. As discussed first in \citet{frag2018}, the NS population always approximately maps the stellar population, in the sense that the ratio of the average NS distance from the cluster centre and the average stellar distance from the cluster centre decreases from nearly unity to $\sim 0.6-0.8$ in a few Gyrs in all the clusters, almost independently of the initial metallicity, binary fraction, and cluster mass. However, the initial cluster metallicity can play a role (shown in color-code in Figure~\ref{fig:rhevol}). BHs in metal-rich clusters are of lower mass and do not inject as much energy into the BH-burning process as more massive BHs in metal-poor clusters \citep{Kremer2019d}. As a consequence, NSs can segregate more efficiently into the dense cluster core, resulting in a smaller NS half-mass radius and a larger spatial overlap with the BH population.

\section{Primordial binary fraction and retention of neutron stars}
\label{sect:binret}

In our models, the initial binary fraction of the OB-type stars, of ZAMS mass $m_{\rm ZAMS}\ge \mcrit$, is assumed to be $\sim 100\,\%$ \citep{baner18}, in order to be consistent with the observed high binary fraction among the massive stars in young clusters and associations \citep[see, e.g.,][]{sanaevans2011,moedist2017}. As described in Section~\ref{sect:models}, most of the our models have $\mcrit=16\msun$. However, the primordial binary fraction and distribution could potentially shape the NS retention and mass spectrum at birth. To preliminarily investigate this, we additionally run cluster models with three different values of the critical mass, $\mcrit=5\msun$, $10\msun$, $16\msun$, above which the primordial binary fraction is assumed to be $\sim 100\,\%$. The initial cluster mass, half-mass radius, and metallicity are fixed to $M_{\rm cl}=1.0\times 10^5\msun$, $r_{\rm h}=1.5$\,pc, and $Z=0.001$, respectively.

\begin{figure} 
\centering
\includegraphics[scale=0.8]{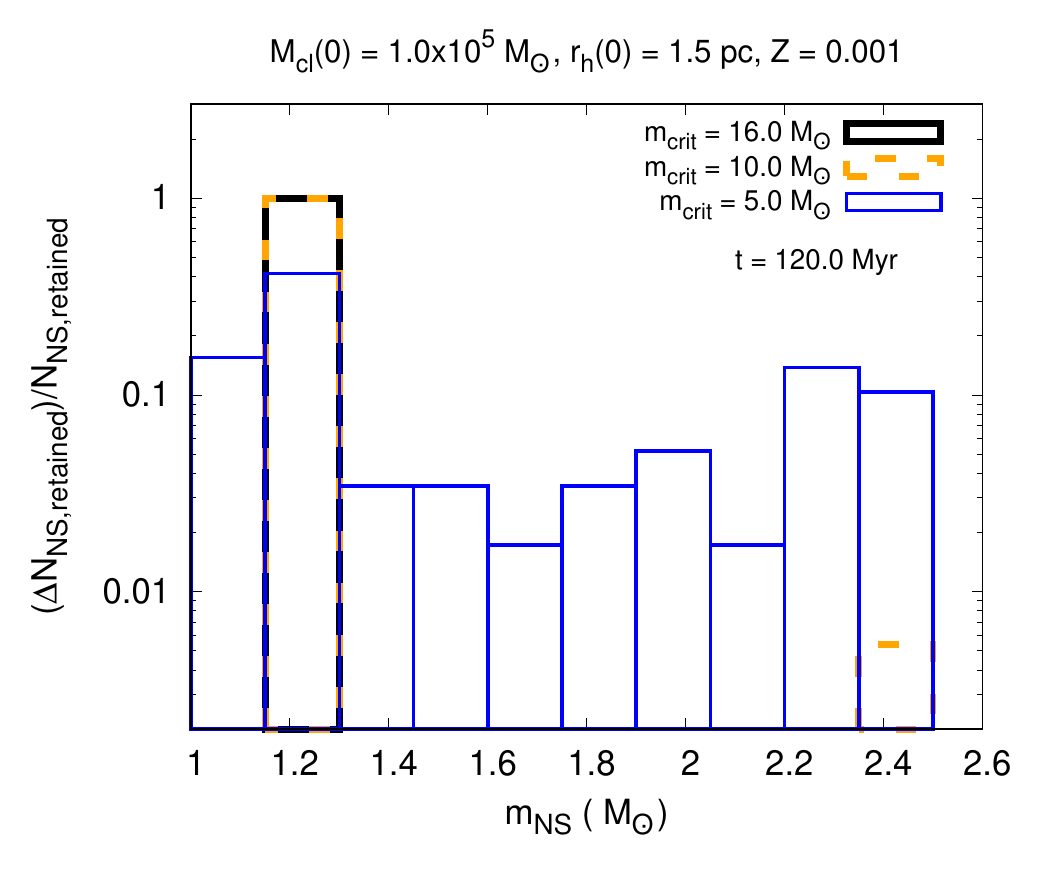}
\includegraphics[scale=0.775]{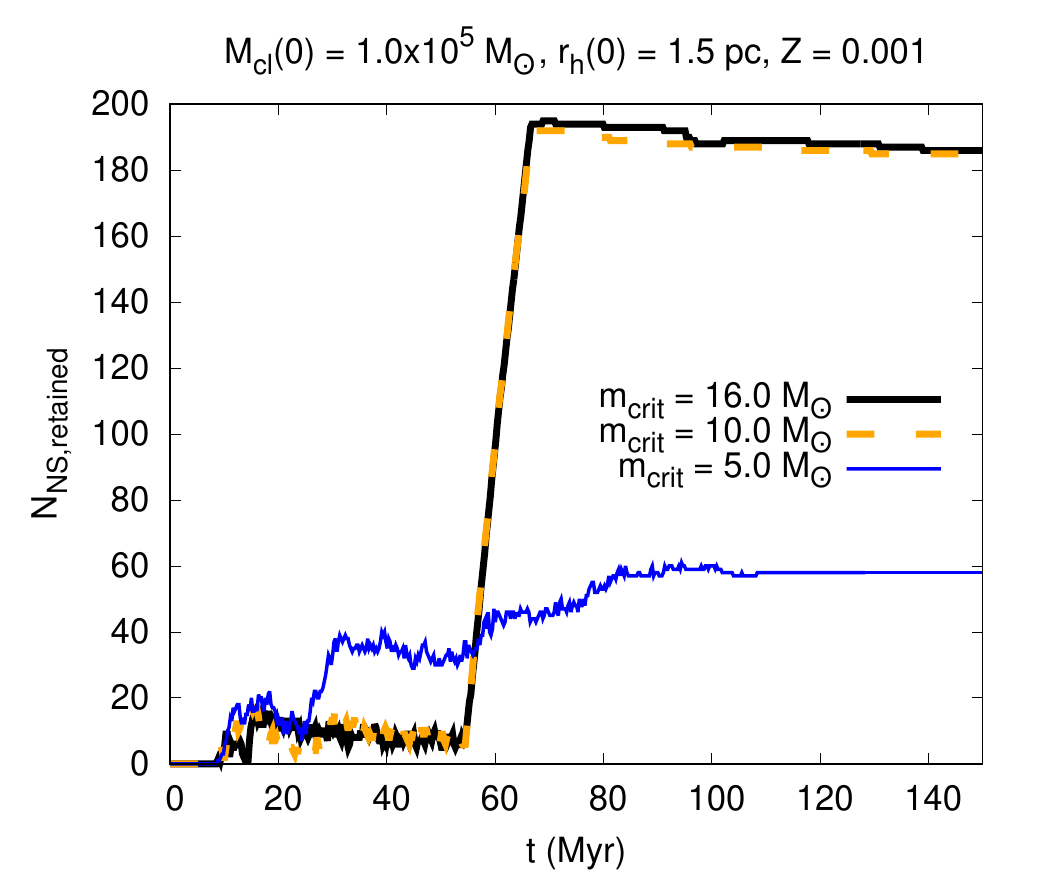}
\caption{Distribution of retained neutron star masses (top panel) and number evolution of retained neutron stars (bottom panel) in a cluster model with initial mass $M_{\rm cl}=1.0\times 10^5\msun$, half-mass radius $r_{\rm h}=1.5$\,pc, and metallicity $Z=0.001$, after all the neutron stars are formed ($t=120$\,Myr). Different colors represent different values of the critical mass, above which the binary fraction of massive stars is taken to be $100\%$: $\mcrit=5\msun$ (blue line), $\mcrit=10\msun$ (yellow line), $\mcrit=16\msun$ (black line).}
\label{fig:mcrit}
\end{figure}

In the top panel of Figure~\ref{fig:mcrit}, we show the mass distribution of NSs retained in the cluster at $120$ Myr \footnote{By this time, NS formation is complete, while dynamics has not significantly shaped NS evolution.}, for different choices of the critical mass. For $\mcrit=10\msun$ and $16\msun$, the majority of the NS-progenitor stars are single (for the chosen $Z=0.001$). Therefore, all the core-collapse NSs would typically receive natal kicks high enough (from a Maxwellian distribution with dispersion of $\sim 265\kms$) to escape the host cluster. Only ECS-NSs receive natal kicks sufficiently low (from a Maxwellian distribution with a dispersion of $\sim 5\kms$) to be retained. As a result, the NSs retained in the cluster at birth will mostly be of $\sim 1.26\msun$ \citep{pod2004}, the mass assumed for ECS-NSs in these models, explaining the roughly delta-function mass distribution of NSs with these values of $\mcrit$ in Figure~\ref{fig:mcrit}.

If the critical mass is fixed to $\mcrit=5\msun$, then, initially, all NS progenitors would have a close binary companion with mass $\ge 5.0\msun$, i.e. most of them will have an NS-progenitor companion (given the assumed standard stellar initial-mass function). Hence, many of the progenitors of ECS-NSs (that are formed latest from a relatively narrow range of progenitor masses) would escape the cluster when the binary they belong disrupts due to the SN mass loss of the companion. In this case, the components of the binary would be released with speeds of the order of the binary's orbital speed \citep{kalogera1996}, which is much larger than cluster escape speed. Whenever the binary is not disrupted, they can also escape along with the binary system. Anyways, the binary center of mass is typically imparted a large velocity boost due to the high natal kick the companion receives when it becomes an NS or a low-mass BH via core-collapse SN. This explains the much reduced peak at $\sim 1.26\msun$ with $\mcrit=5\msun$ in Figure~\ref{fig:mcrit}.

Whatever the initial value of the critical mass is, the $\sim 1.26\msun$ ECS-NSs still comprise the dominant NS population. Therefore, the overall buildup with time of the NS population in the cluster is inhibited for $\mcrit=5\msun$, as shown in the bottom panel of Figure~\ref{fig:mcrit}. We note that binary mergers (due to binary evolution and dynamical interactions) leading to single-NS or BH progenitors also contribute to building a smaller NS population. The ionization of a binary, driven primarily by the natal kick of the newly formed core-collapse-SN NS, may also cause the latter to slow down sufficiently (due to the kinetic energy absorption in snapping the binary), so that it is retained in the cluster, which would have escaped if born as a single. As a result, some core-collapse-SN NSs stay back in the cluster, giving rise to the tails around $\sim 1.26\msun$ in the NS mass distribution in Figure~\ref{fig:mcrit}, when $\mcrit=5\msun$. Note that dynamical ionization is also, in principle, possible for all binaries containing NS or NS progenitors. However, given that the majority of the NS progenitors are in tight, hard binaries, dynamical ionization of such binaries is generally inefficient in the present models.

Note that the tail of the NS mass distribution exceeds the upper mass limit of about $2\msun$ for NSs in the rapid remnant-mass prescription \citep[e.g.][]{baner2019bse}. These few NSs have increased their mass through mass accretion from their binary companions after their formation. We also note that the outcomes of binary evolution inside a cluster can be significantly influenced by dynamical encounters and can be irreproducible in isolated evolution of the same set of initial binaries\footnote{The tail could be affected by the specific ways the binary-evolution engine and the natal kicks are handled in symbiotic and mass-transferring binaries in \textsc{Nbody7/BSE}.}.

\section{Orbital properties of binaries comprised of neutron stars}
\label{sect:nsbinary}

In star clusters, NSs can form binaries either through stellar evolution of primordial binaries or as a result of dynamical exchange interactions. Even if the primordial membership is maintained, the orbital parameters can significantly be altered via dynamical encounters in dense star clusters, so they are not direct outcomes of isolated binary evolution. As discussed in Section~\ref{sect:masseg}, NSs are able to segregate in the innermost and densest cluster regions, where they can efficiently form NS-NS and BH-NS binaries, only when the BH population is depleted enough and the BH-burning process is weakened.

\begin{figure} 
\centering
\includegraphics[scale=0.565]{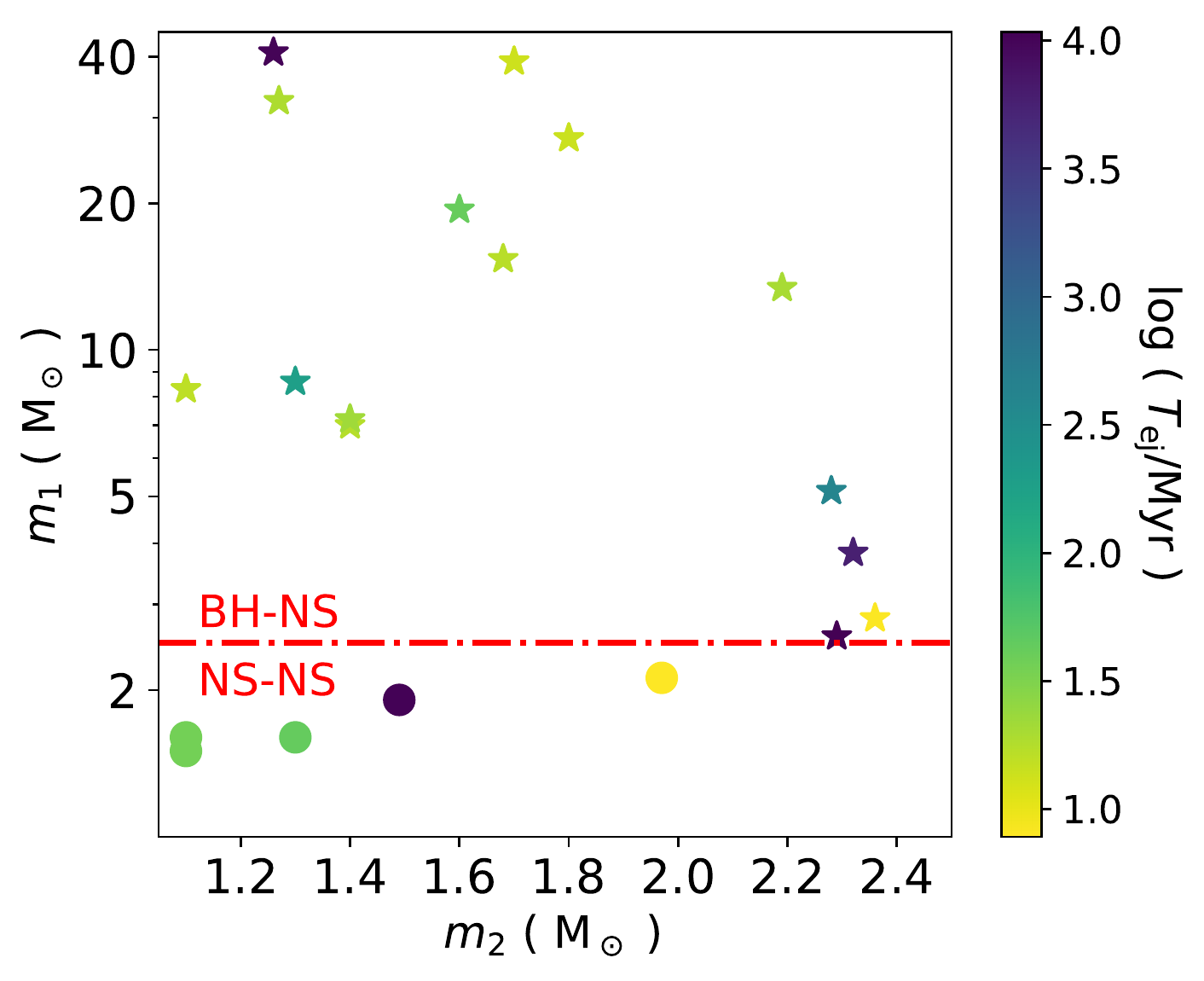}
\includegraphics[scale=0.565]{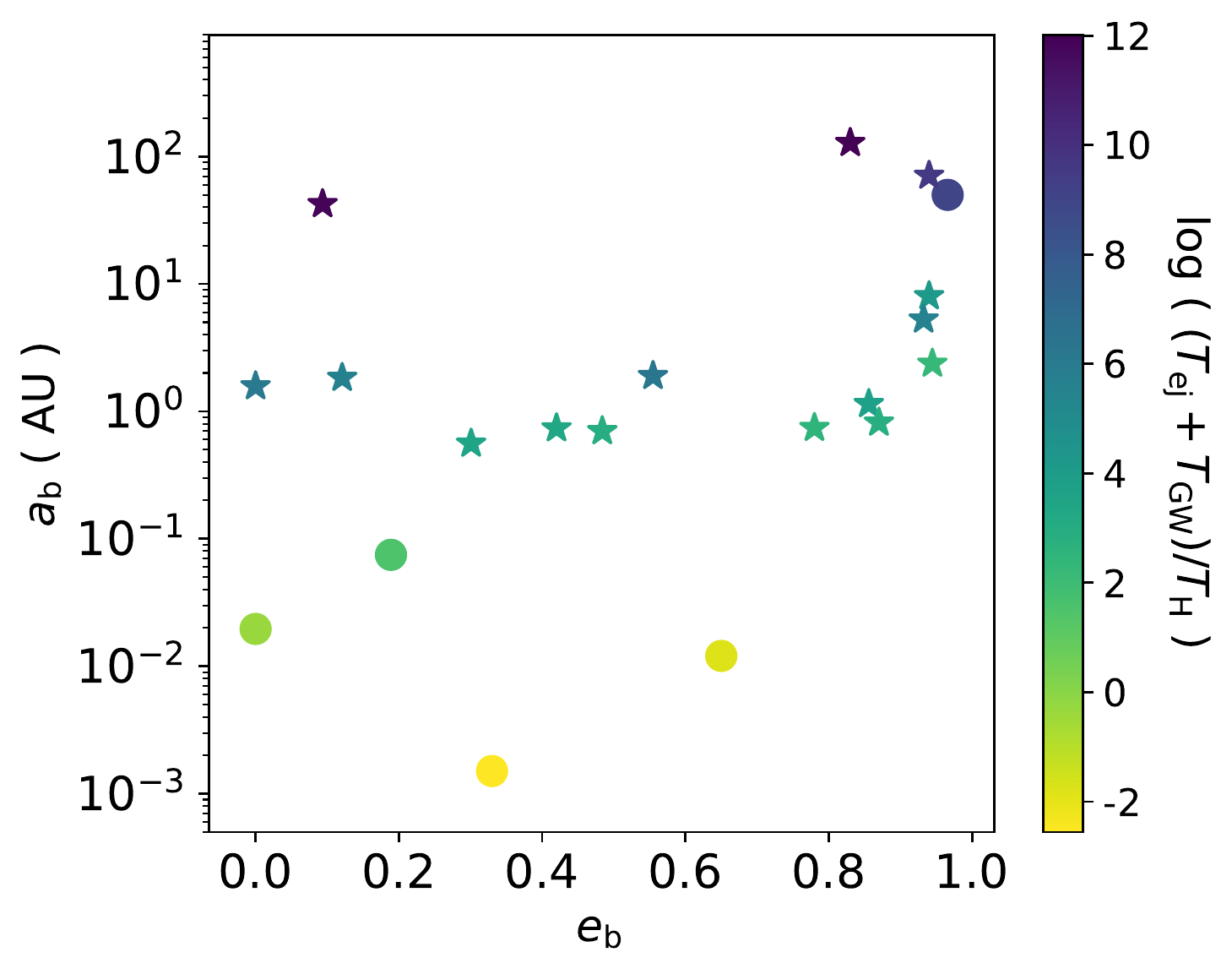}
\caption{Masses ($m_1$-$m_2$; top panel) and the orbital semi-major axis and eccentricity ($a_{\rm b}$-$e_{\rm b}$; bottom panel) of the NS-NS (circles) and BH-NS (stars) binaries ejected from the cluster models presented in \citet{banerjee2020}. Color code: merger time ($T_{\rm ej}+T_{\rm GW}$) in units of the Hubble time.}
\label{fig:binmassmerg}
\end{figure}

We show in Figure~\ref{fig:binmassmerg} the masses ($m_1$-$m_2$; top panel) and the orbital semi-major axis and eccentricity ($a_{\rm b}$-$e_{\rm b}$; bottom panel) of the NS-NS and BH-NS binaries ejected from the cluster models presented in \citet{banerjee2020}. There are $20$ ejected NS-NS and BH-NS binaries in total. We find that $80\%$ of these ejected binaries, especially those with early ejection times, maintain the primordial membership, while only $20\%$ of them are exchanged systems. The cause of ejection can be either a natal kick or a close encounter with other cluster members. Clusters with a smaller half-mass radius would dynamically process and deplete BHs faster, enabling segregation and interaction among lighter members earlier \citep[e.g.,][]{kremer2019initial}. Hence, as expected, we find that the exchanged binaries are formed only in clusters with the smallest half-mass radii and ejected at late evolutionary times ($\gtrsim 1$ Gyr). In contrast, the binaries that maintain their primordial membership are ejected much earlier (within $\lesssim200{\rm~Myr}$) and from also the less dense models. Also, the exchanged NS-NS/BH-NS binaries are always ejected dynamically while only a few of those preserving the primordial membership are dynamically ejected. In total, $30$\% of the NS-NS/BH-NS binaries are dynamically ejected (after $\gtrsim100{\rm~Myr}$) and $70$\% are ejected due to natal kick (within $\lesssim50{\rm~Myr}$). As shown in the top panel of Figure~\ref{fig:binmassmerg}, we produce $4$ BH-NS systems, where the BH is in the nominal BH low-mass gap ($\sim 3$--$5\msun$). This is due to the fact that the delayed-SN prescription \citep{fryer2012} is adopted in some of our models.

We show in Figure~\ref{fig:binmassmerg} in color code the merger time in units of the Hubble time ($T_{\rm H}$). The merger time is computed as the sum of the ejection time from the cluster ($T_{\rm ej}$) and the GW merger time \citep{peters1964gravitational}
\begin{equation}
T_{\rm GW}=\frac{5}{256}\frac{a_{\rm b}^4 c^5}{G^3 m_1 m_2(m_1+m_2)}(1-e_{\rm b}^2)^{7/2}\ .
\label{eqn:tgw12}
\end{equation}
We find that none of the BH-NS binaries merges, while $3$ out of $5$ NS-NS binaries merge within a Hubble time. All of them come from systems that maintain their primordial membership. Assuming a cluster density of $\sim 2.31\,\rm{Mpc}^{-3}$ \citep{Rodriguez2015a} and that $3$ NS-NS merge within a Hubble time among all our cluster models. The NS-NS binaries that merge come from the runs with $m_{\rm crit}=5\msun$. We estimate a rate of $\sim 0.01\gpcyr$ for NS-NS mergers in the local Universe. We also compute an upper limit of $\sim 3\times 10^{-3}\gpcyr$ for the BH-NS merger rate. We note that the local density of the latter could be $\sim 10$ times higher \citep{Banerjee2017stellara,baner18}, and our predicted NS-NS merger rate density could be as high as $\sim 0.1\gpcyr$, with an upper limit of $\sim 3\times 10^{-2}\gpcyr$ for the BH-NS merger rate. In contrast to \citet{rastello2020} and \citet{santoliquido2020}, our inferred rate is several orders of magnitude smaller than the current empirical merger rate from LIGO/Virgo. Thus our results are well in agreement with previous estimates on the merging NS-NS and BH-NS binaries from the cluster dynamical channel \citep{bae2014compact,clausen2013black,belczynski2018origin,ye2020}. 

The inference of much higher BH-NS and NS-NS merger rate densities from young clusters in \citet{rastello2020} and \citet{santoliquido2020} is likely a result of the combination of the initial fractal structure, predominantly $\sim 0.2$ pc initial sizes, and mostly $N\sim10^2-10^3$ in the model clusters employed in these studies which features all enable early and efficient mass segregation. Furthermore, all stars are assumed to form in such clusters without any short-term dissolution mechanism (e.g., residual gas dispersal; \citealt{bankroupa2018}) for low mass clusters, over a large cosmic-evolutionary time (i.e. effectively assuming zero infant mortality of all clusters), leading to optimistic merger rates. In contrast, our models' initial configurations are more dispersed, with $\sim$ pc sizes, and monolithic and they are one to three orders of magnitude more massive, leading to significantly longer two body-relaxation and hence mass-segregation times. However, they are consistent with the observed structure and kinematics of gas-free young massive clusters and moderately massive open clusters \citep{portgz2010}. Furthermore, we estimate the merger rate density based directly on observed local spatial density
of clusters (and scaling that for young clusters), while \citet{rastello2020} and \citet{santoliquido2020} assume that all star formation goes into small clusters. 

\section{Discussion and conclusions}
\label{sect:conc}

NS-NS and BH-NS binaries are among the most interesting astrophysical objects, being precursors of GW events and gamma-ray bursts. Dense star clusters are the natural environment where hundreds of NSs can form and dynamically evolve. Owing to high stellar densities and low velocity dispersions, NSs can interact with single and binary stars to form bound systems, which can, potentially, later merge and be observed via GW emission. So far, two NS-NS binaries have been observed by the LIGO/Virgo collaboration, with an inferred merger rate in the local Universe of $\sim 250$--$2810\gpcyr$ \citep{abbott2017gw170817,abb2020}. No BH-NS binaries have been confirmed, with a LIGO/Virgo $90\%$ upper limit of $\sim 610\gpcyr$ on the merger rate \citep{abb2019cat}. Note, although, that GW190814 can, possibly, be a BH-NS merger \citep{abb2020}.

In this paper, we have studied the origin and dynamical evolution of NSs within clusters with different initial masses, metallicities, primordial binary fractions, and prescriptions for the natal kicks imparted to NSs and BHs at birth. We have found that the radial profile of NSs is shaped by the BH content of the cluster, which partially quenches the NS segregation due to the BH-burning process. These leaves most of the NSs out of the densest cluster regions, where NS-NS and BH-NS binaries can be formed. Due to a large velocity kick that they receive at birth, most of the NSs escape the host clusters, with the bulk of their retained population made up of NSs of $\sim 1.3\msun$, coming from the electron-capture supernova process. The primordial binary fraction and pairing can smear out this distribution. Such inferences would be of interest for pulsar searches in young and open clusters.

Finally, we have found no BH-NS mergers, while we find that a subset of our models produce NS-NS mergers, leading to a rate of $\sim 0.01\gpcyr$ in the local Universe, several orders of magnitude smaller than the current empirical merger rate from LIGO/Virgo. Therefore, in order to account for the observed LIGO/Virgo rates, other channels or environments need to be invoked, such as binary evolution with varying common envelope parameters and natal kicks \citep{kruc2018,Baibhav_2019}, triple star scenarios \citep{frl2019a,frl2019b,hamersthm2019} in the field, or low-mass young star clusters \citep{rastello2020,DiCarlo_2020,santoliquido2020}.

\section*{Acknowledgements}

We thank the anonymous referee for constructive comments and suggestions that have helped to improve the manuscript.
We thank Fred Rasio, Kyle Kremer, Claire Shi Ye, and John Antoniadis for useful discussions. GF acknowledges support from a CIERA postdoctoral fellowship at Northwestern University. SB acknowledges the support from the Deutsche Forschungsgemeinschaft (DFG; German Research Foundation) through the individual research grant "The dynamics of stellar mass black holes in dense stellar systems and their role in gravitational-wave generation" (BA 4281/6-1; PI: S. Banerjee). SB acknowledges the generous support and efficient system maintenance of the computing teams at the AIfA and HISKP.

\bibliographystyle{yahapj}
\bibliography{refs}

\end{document}